\documentclass[12pt]{article}
\usepackage{sc3conf}
\usepackage{amsfonts}
\usepackage{epsfig}
\def\appendix{\par\clearpage
  \setcounter{section}{0}
  \setcounter{subsection}{0}
  \@addtoreset{equation}{section}
  \def\@sectname{Appendix~}
  \def\theequation{\thesection\arabic{equation}}
  \def\thesection{\Alph{section}}}
 
\def\thefigures#1{\par\clearpage\section*{Figures\@mkboth
  {FIGURES}{FIGURES}}\list
  {Fig.~\arabic{enumi}.}{\labelwidth\parindent\advance
\labelwidth -\labelsep
      \leftmargin\parindent\usecounter{enumi}}}

\def\thetables#1{\par\clearpage\section*{Tables\@mkboth
  {TABLES}{TABLES}}\list
  {Table~\Roman{enumi}.}{\labelwidth-\labelsep
      \leftmargin0pt\usecounter{enumi}}}

\def\@sect#1#2#3#4#5#6[#7]#8{\ifnum #2>\c@secnumdepth
     \def\@svsec{}\else
     \refstepcounter{#1}\edef\@svsec{\@sectname\csname the#1\endcsname
.\hskip 1em }\fi
     \@tempskipa #5\relax
      \ifdim \@tempskipa>\z@
        \begingroup #6\relax
          \@hangfrom{\hskip #3\relax\@svsec}{\interlinepenalty \@M #8\par}
        \endgroup
       \csname #1mark\endcsname{#7}\addcontentsline
         {toc}{#1}{\ifnum #2>\c@secnumdepth \else
                      \protect\numberline{\csname the#1\endcsname}\fi
                    #7}\else
        \def\@svse=chd{#6\hskip #3\@svsec #8\csname #1mark\endcsname
                      {#7}\addcontentsline
                           {toc}{#1}{\ifnum #2>\c@secnumdepth \else
                             \protect\numberline{\csname the#1\endcsname}\fi
                       #7}}\fi
     \@xsect{#5}}
 
\def\@sectname{}
\newcommand{\be}{\begin{equation}}
\newcommand{\ee}{\end{equation}\noindent}
\newcommand{\bear}{\begin{eqnarray}}
\newcommand{\ear}{\end{eqnarray}\noindent}
\newcommand{\benn}{\begin{enumerate}}
\newcommand{\enn}{\end{enumerate}}
\newcommand{\no}{\noindent}
\date{}
\renewcommand{\theequation}{\arabic{section}.\arabic{equation}}

\def\Eins{\mathord{1\hskip -1.5pt
\vrule width .5pt height 7.75pt depth -.2pt \hskip -1.2pt
\vrule width 2.5pt height .3pt depth -.05pt \hskip 1.5pt}}

\newcommand{\slD}{\raise.15ex\hbox{$/$}\kern-.57em\hbox{$D$}}
\newcommand{\slpartial}{\raise.15ex\hbox{$/$}\kern-.57em\hbox{$\partial$}}
\newcommand{\slG}{{{\dot G}\!\!\!\! \raise.15ex\hbox {/}}}

\def\GBd12{{\dot G}_{B12}}

\def\non{\nonumber}
\def\beqn*{\begin{eqnarray*}}
\def\eqn*{\end{eqnarray*}}

\def\square{\kern1pt\vbox{\hrule height 1.2pt\hbox{\vrule width 1.2pt
   \hskip 3pt\vbox{\vskip 6pt}\hskip 3pt\vrule width 0.6pt}
   \hrule height 0.6pt}\kern1pt}

\def\slash#1{#1\!\!\!\raise.15ex\hbox {/}}

\def\dps{\displaystyle}

\def\half{{1\over 2}}

\def\e{\mbox{e}}

\def\4piTD{{(4\pi T)}^{-{D\over 2}}}
\def\4piT4{{(4\pi T)}^{-2}}

\def\Tintm4{{\dps\int_{0}^{\infty}}{dT\over T}\,e^{-m^2T}
    {(4\pi T)}^{-2}}
\def\Tintm{{\dps\int_{0}^{\infty}}{dT\over T}\,e^{-m^2T}}

\def\mn{{\mu\nu}}

\def\bbbz{{\mathchoice {\hbox{$\sf\textstyle Z\kern-0.4em Z$}}
{\hbox{$\sf\textstyle Z\kern-0.4em Z$}}
{\hbox{$\sf\scriptstyle Z\kern-0.3em Z$}}
{\hbox{$\sf\scriptscriptstyle Z\kern-0.2em Z$}}}}

\begin{document}
\pagestyle{plain}
\raggedbottom

\hfill UMSNH-PHYS-02-11
\vspace{17pt}
\title{Two-loop self-dual QED}

\authors{Gerald V. Dunne,\adref{1}
  \underline{Christian Schubert}\adref{2}}

\addresses{\1ad Department of Physics, University of Connecticut,
Storrs, CT 06269-3046, USA
  \nextaddress \2ad Instituto de F\'{\i}sica y Matem\'aticas,
Universidad Michoacana de San Nicol\'as de Hidalgo,
Apdo. Postal 2-82,
C.P. 58040, Morelia, Michoac\'an, M\'exico.}

\maketitle
\vspace{6pt}
\centerline{Talk given by C. S. at {\sl 3rd International Sakharov}}
\centerline{{\sl Conference On Physics}, 24-29 June 2002, Moscow.}
\vspace{-8pt}

\begin{abstract}
We present explicit closed-form expressions for the two-loop
Euler-Heisenberg Lagrangians in a constant self-dual field,
for both spinor and scalar QED. The simplicity of these
representations allows us to examine in detail the asymptotic 
properties of these Lagrangians, and to construct their imaginary
part using Borel dispersion relations. In particular, for this
self-dual case we obtain the explicit form of the Lebedev-Ritus
functions appearing in the Schwinger representation of the
imaginary part at two loops. Using the connection between
self-duality and helicity, we also obtain explicit formulas
for the low energy limits of the `all + helicity' 
$N$ photon amplitudes, in scalar and spinor QED at one and two loops.  
\end{abstract}

\vspace{-5pt}
\section{Introduction}

In gauge theory a very special role is played by self-dual
fields, i.e. fields satisfying the condition

\vspace{-12pt}\bear
F_{\mu\nu}=\tilde
F_{\mu\nu}\equiv\frac{1}{2}\varepsilon_{\mu\nu\alpha\beta}
F^{\alpha\beta}
\label{sdcond}
\ear
Fields of this type are prominent in QCD for a number of different
reasons:

\begin{itemize}

\item
Instantons are self-dual.

\item
Among all covariantly constant gluon backgrounds, only the self-dual
quasi-abelian background is stable (at one-loop) under fluctuations
\cite{leutwyler}.

\item
Large classes of integrable models can be obtained by dimensional
reduction starting from self-dual Yang-Mills theories 
\cite{ward}.

\item
Self-dual fields are helicity eigenstates, so that the effective
action in such a field carries the information on the corresponding
gluon amplitudes with all equal helicities \cite{dufish}. Such `all $+$'
amplitudes generally exhibit a particularly simple structure,
at the tree level \cite{allplustree} and beyond \cite{allplusloop}.

\end{itemize}
In the abelian case most of this motivation does not exist, with the
exception of the last point mentioned, which holds in the abelian
case as well. Correspondingly, little use has been made so far 
of self-dual fields in QED. In the present contribution, we will
consider the case of {\sl constant} self-dual fields in QED, and
show that the self-duality condition leads to very remarkable 
simplifications for the effective action in such a background field
\cite{ds}. 

\section{One-loop Euler-Heisenberg Lagrangians}

Before presenting these two-loop results, let us shortly recapitulate
some facts about Euler-Heisenberg Lagrangians.
At one-loop, the 
on-shell renormalized effective Lagrangians in a constant background
field, for spinor and
scalar QED, are given by the well-known formulas \cite{eh,schwinger}

\bear
{\cal L}_{\rm spin}^{(1)} 
&=&
-
{1\over 8\pi^2}
\int_0^{\infty}{dT\over T}
\,\e^{-m^2T}
\biggl[
{e^2ab\over \tanh(eaT)\tan(ebT)}
-{e^2\over 3} (a^2-b^2) -{1\over T^2}
\biggr]
\non\\
{\cal L}_{\rm scal}^{(1)} 
&=&
{1\over 16\pi^2}
\int_0^{\infty}{dT\over T}
\,\e^{-m^2T}
\biggl[
{e^2ab\over \sinh(eaT)\sin(ebT)}
+{e^2\over 6} (a^2-b^2) -{1\over T^2}
\biggr]
\non\\
\label{L1}
\ear
Here $a,b$ are related to the two
invariants of the Maxwell field by
$a^2-b^2=B^2-E^2, ab = {\bf E}\cdot{\bf B}$.

These effective Lagrangians are
real for a purely magnetic field, while in the presence of an electric 
field there is
an imaginary (absorptive) part, indicating the process of 
electron--positron (resp.
scalar--antiscalar) pair creation by the field. 
For example, in the case of a purely electric field, $E$, 
the effective Lagrangians 
(\ref{L1}) have imaginary parts given by:
\begin{eqnarray}
{\rm Im} {\cal L}_{\rm spin}^{(1)}(E) &=&  \frac{m^4}{8\pi^3}
\beta^2\, \sum_{k=1}^\infty \frac{1}{k^2}
\,\exp\left[-\frac{\pi k}{\beta}\right]
\non\\
{\rm Im}{\cal L}_{\rm scal}^{(1)}(E) 
&=&
-\frac{m^4}{16\pi^3}
\beta^2\, \sum_{k=1}^\infty \frac{(-1)^{k}}{k^2}
\,\exp\left[-\frac{\pi k}{\beta}\right]
\label{L1im}
\end{eqnarray}
where $\beta = eE/m^2$. These expressions are clearly non-perturbative in 
terms of
the field and coupling.
Their physical interpretation
is that the coefficient of the $k$-th
exponential can be directly identified with the rate for the coherent 
production of
$k$ pairs by the field \cite{schwinger}.

\section{Two-loop Euler-Heisenberg Lagrangians}

The two-loop corrections to the Euler-Heisenberg
Lagrangians, taking into account an
additional photon exchange in the loop, where first calculated by Ritus
in the seventies \cite{ritus} (see also
\cite{ditreu}). 
More recently these Lagrangians have been recalculated 
\cite{frss} using the `string-inspired' formalism 
\cite{berkos,strassler,ss1,review}.
However, in all cases the
results are not nearly as explicit as the one-loop formulas 
(\ref{L1}); they involve
two-parameter integrals and a counterterm from one-loop mass renormalization.
As far as the magnetic case is concerned, where the
effective Lagrangian is real, the complicated form of
these representations is perhaps not too bothersome.
For values of the magnetic field small compared to
the `critical' field strength $B_c \equiv {m^2\over e}$,
the Lagrangian can be computed using the weak field expansion,
whose coefficients are easy to compute
to fairly high orders \cite{frss,dunsch}:

\bear
{\cal L}_{\rm spin}^{(2)}[B]
&=&
{\alpha m^4\over
{(4\pi )}^{3}}
{1\over 81}
\Biggl[
64 
{\Bigl({B\over B_{\rm cr}}\Bigr)}^4
-{1219\over 25}
{\Bigl({B\over B_{\rm cr}}\Bigr)}^6
+ {135308\over 1225}
{\Bigl({B\over B_{\rm cr}}\Bigr)}^8
- \ldots \, 
\Biggr]
\non\\
\label{wfe}
\ear
For general values of the field strength numerical integration
can be used.

Calculating the imaginary part of the corresponding
Lagrangian for the electric field case is a more difficult matter.
At the one-loop level, the imaginary parts (\ref{L1im})
can be obtained from (\ref{L1}) 
by a simple application of the
residue theorem. The analogous analysis for the two-loop
parameter integrals is already highly nontrivial. Nevertheless,
Ritus and Lebedev \cite{ritus,lebrit} were able
to obtain along these lines the following two-loop generalization
of the Schwinger decompositions (\ref{L1im}):

\begin{eqnarray}
{\rm Im} {\cal L}_{\rm spin}^{(1)} (E) +
{\rm Im} {\cal L}_{\rm spin}^{(2)} (E)
&=&  \frac{m^4}{8\pi^3}
\beta^2\,
\sum_{k=1}^\infty
\Bigl[
\frac{1}{k^2}
+\alpha\pi K_k^{\rm spin}(\beta)
\Bigr]
\,\exp\left[-\frac{\pi k}{\beta}\right]
\non\\
{\rm Im} {\cal L}_{\rm scal}^{(1)} (E) +
{\rm Im} {\cal L}_{\rm scal}^{(2)} (E)
&=&  -\frac{m^4}{16\pi^3}
\beta^2\,
\sum_{k=1}^\infty
(-1)^k
\Bigl[
\frac{1}{k^2}
+\alpha\pi K_k^{\rm scal}(\beta)
\Bigr]
\,\exp\left[-\frac{\pi k}{\beta}\right]
\non\\
\label{fullimag2l}
\end{eqnarray}
where $\alpha=\frac{e^2}{4\pi}$ is the fine-structure constant. 
The coefficient
functions $K_k^{\rm spin, scal}(\beta)$ appearing here were not 
obtained explicitly by \cite{lebrit}.
However, it was shown that they 
have small $\beta$ expansions of the following form:

\begin{eqnarray}
K_k^{\rm spin,scal}(\beta) 
&=& -{c_k\over \sqrt{\beta}} + 1 + {\rm O}(\sqrt{\beta})
 \nonumber\\
c_1 = 0,\quad && \quad
c_k = {1\over 2\sqrt{k}}
\sum_{l=1}^{k-1} {1\over \sqrt{l(k-l)}},
\quad k \geq 2
\label{expK}
\end{eqnarray}
Note that to the order given 
$K_k^{\rm spin}=K_k^{\rm scal}$.
For $k\geq 2$, these expansions start with terms that
are singular in the limit of vanishing field $\beta\to 0$,
which seems to be at variance with the fact that these
coefficients have a direct physical meaning.
In \cite{lebrit} a physically intuitive solution
was offered to this dilemma. Its basic assumption is
that, if one would take into account 
contributions from higher loop orders to the prefactor
of the $k$-th exponential, then one would find the
two lowest order terms 
in the
small -- $\beta$ expansion of $K_k^{\rm spin,scal}(\beta)$
to exponentiate in the following way,

\begin{eqnarray}
\Bigl[
\frac{1}{k^2}
+\alpha\pi K_k^{\rm spin,scal}
\bigl({eE\over m^2}\bigr)
+\ldots
\Bigr]
\,\exp\left[-{k\pi m^2\over eE}\right]
=
\frac{1}{k^2}
\exp\left[-{k\pi m^2_{\ast}(k,E)\over eE}\right]
\non\\
\label{shiftm}
\end{eqnarray}
It would thus be possible to absorb their effect
completely into
a field-dependent shift of the electron mass:

\bear
m_{\ast}(k,E) &=&
m +\half\alpha kc_k\sqrt{eE}-\half\alpha keE/m
\label{massshift}
\ear
Moreover, these
contributions to the mass shift have a simple meaning in
the coherent tunneling picture
\cite{lebrit}:
The negative term can be interpreted as the total 
Coulomb energy of
attraction between opposite charges in
a coherent group; the positive term,
which is present only in the case $k\geq 2$, represents the
energy of repulsion between like charges.
For the attractive term, a completely
different derivation of the same exponentiation
was given in \cite{afalma}. 

\section{The imaginary part via Borel dispersion
relations}

Clearly, it would be of interest to understand in greater detail 
this prefactor series
$K_k^{\rm spin,scal}(\beta)$. 
In an effort to learn more about it,
in \cite{dunsch} we used Borel techniques to study the 
imaginary part of
the two-loop effective Lagrangian for a constant electric field background. 

Let us review some basic facts on Borel summation.
Consider an asymptotic series expansion of
some function
$f(g)$
\begin{eqnarray}
f(g)\sim \sum_{n=0}^\infty \, a_n\, g^n
\label{exp}
\end{eqnarray}
where $g\to 0^+$ is a small dimensionless perturbation expansion parameter.
In many physics applications 
perturbation theory leads 
to a divergent
series in which the expansion coefficients 
$a_n$ have leading large-order
behaviour

\begin{eqnarray}
a_n\sim (-1)^n \rho^n \Gamma(\mu\, n+\nu)  \qquad\qquad (n\to\infty)
\label{general}
\end{eqnarray}
for some real constants $\rho$, $\mu>0$, and $\nu$. When $\rho>0$, the
perturbative expansion coefficients $a_n$ alternate in sign and their
magnitude grows factorially, just as in the Euler-Heisenberg case.
Borel summation is a useful approach to this case of a
divergent, but alternating series. 
The leading Borel
approximation is 

\begin{eqnarray}
f(g)\sim \frac{1}{\mu}\, \int_0^\infty \frac{ds}{s} \,
\left(\frac{1}{1+s}\right)
\left(\frac{s}{\rho g}\right)^{\nu/\mu}\,
\exp\left[-\left(\frac{s}{\rho g}\right)^{1/\mu}\right]
\label{genborel}
\end{eqnarray}
For a non-alternating series, we need $f(-g)$. The
Borel integral (\ref{genborel}) is an analytic function of $g$
in the cut $g$ plane: $|{\rm arg}(g)|<\pi$. So a dispersion relation (using
the discontinuity across the cut along the negative $g$ axis) can be used to
{\it define} the imaginary part of $f(g)$ for negative values of the expansion
parameter:

\begin{eqnarray}
{\rm Im} f(-g)\sim\frac{\pi}{\mu}\left(\frac{1}{\rho g} \right)^{\nu/\mu}
\exp\left[-\left(\frac{1}{\rho g}\right)^{1/\mu}\right]
\label{genimag}
\end{eqnarray}

Returning to the Euler-Heisenberg Lagrangian, 
for a uniform magnetic background the weak-field
expansions of the one-loop Lagrangians (\ref{L1}) 
are precisely of the form (\ref{exp}),(\ref{general}) with
$g=(\frac{eB}{m^2})^2$. For example, in the spinor QED case one has

\begin{eqnarray}
a_n^{(1)}&=&-\frac{2^{2n}{\cal B}_{2n+4}} {(2n+4)(2n+3)(2n+2)}
\sim (-1)^{n}\frac{1}{8 \pi^4}\,\frac{\Gamma(2n+2)}{\pi^{2n}}\left(
1+\frac{1}{2^{2n+4}}+\dots\right)
\non\\
\label{growth}
\end{eqnarray}
where the ${\cal B}_m$ are Bernoulli numbers.
For a
uniform {\it electric} background, the only difference perturbatively is that
$B^2$ is replaced by $-E^2$; that is, $g=(\frac{eB}{m^2})^2$ is replaced by
$-g=-(\frac{eE}{m^2})^2$. So the perturbative one-loop Euler-Heisenberg
series becomes non-alternating. Then from (\ref{genimag}), with
$\rho=\frac{1}{\pi^2}$ and $\mu=\nu=2$, we immediately deduce the leading
behaviour of the imaginary part of the one-loop Euler-Heisenberg effective
Lagrangian:
\begin{eqnarray}
{\rm Im} {\cal L}^{(1)}_{\rm spin}(E) \sim
\frac{m^4}{8\pi^3}\left(\frac{eE}{m^2}\right)^2\,
\exp\left[-\frac{m^2\pi}{eE}\right]
\label{leadingimag}
\end{eqnarray}
Taking into account also the
sub-leading corrections (\ref{growth}) to the leading large-order behaviour of
the expansion coefficients $a_n^{(1)}$, one can apply (\ref{genimag})
successively to reconstruct the full Schwinger series
(\ref{L1im}).

At the two-loop level, no closed-form expression is known for the
corresponding expansion coefficients $a_n^{(2)}$, so that the
corresponding analysis becomes much more involved.
In \cite{dunsch}, first 
the integral representation for the purely magnetic case obtained in  
\cite{frss} was 
used to compute these coefficients up to $n=15$.
It was then established by a numerical analysis that

\begin{eqnarray}
a_n^{(2)}\sim (-1)^n \frac{16}{\pi^2} \, \frac{\Gamma(2n+2)}{\pi^{2n}}\left[
1-\frac{0.44}{\sqrt{n}}+\dots \right]
\label{2lcorr}
\end{eqnarray}
Thus the leading large-order growth corresponds to the form (\ref{general}),
and moreover differs from the one-loop case (\ref{growth}) only by
a global prefactor. To the contrary, the subleading correction term
given in (\ref{2lcorr}) shows
a much weaker $n$ dependence than is found for the first
correction in the one-loop case (\ref{growth}). This means that in the
two-loop case the dominant corrections are to the prefactor in the leading
behaviour. This is in contrast to the one-loop case
where the first correction to the leading behaviour is
exponentially suppressed. Indeed, applying the Borel relations, the correction
term (\ref{2lcorr}) leads to
\begin{eqnarray}
{\rm Im} \left({\cal L}^{(1)}_{\rm spin}(E)
+{\cal L}^{(2)}_{\rm spin}(E)\right) &\sim&
\left(1+\alpha\,\pi\left[1- (0.44) \sqrt{\frac{2eE}{\pi
m^2}}+\dots\right]\right)
\non\\&&\times
\frac{m^4}{8\pi^3}\left(\frac{eE}{m^2}\right)^2\,
\exp\left[-\frac{m^2\pi}{eE}\right]
\label{nextsum}
\end{eqnarray}
Thus the structure of
(\ref{nextsum}) conforms already to the form 
(\ref{fullimag2l}),(\ref{expK}).

\section{The self-dual case}

It seems very difficult to make further progress along these lines
for the purely magnetic/electric field cases, 
let alone the general constant field
case. To the contrary, in the self-dual case this analysis can be
carried much further \cite{ds}. Technically, this is
because the self-duality condition (\ref{sdcond}) implies that

\bear
F^2 &=& -f^2\Eins
\label{deff}
\ear
where $f^2 = {1\over 4}F_{\mn}F^{\mn}$, and $\Eins$ denotes the
identity matrix in Lorentz space. This leads to enormous simplifications
for this type of calculations.

In the self-dual (`SD') case, the integral representation (\ref{L1})
for the renormalized one-loop scalar effective Lagrangian becomes

\begin{eqnarray}
{\cal L}_{\rm
scal}^{(1)(SD)}(\kappa)=\frac{m^4}{(4\pi)^2}\frac{1}{4\kappa^2}\int_0^\infty
\frac{dt}{t^3}\, e^{-2\kappa t}\left[\frac{t^2}{\sinh^2(t)}-1+\frac{t^2}{3}\right]
\label{1lsclag}
\end{eqnarray}
where $\kappa\equiv {m^2\over 2e\sqrt{f^2}}$ is the natural 
dimensionless parameter.
The Lagrangian for the spinor QED case differs from this only by the standard
global factor of $-2$ for statistics and degrees of freedom:

\begin{eqnarray}
{\cal L}_{\rm spin}^{(1)(SD)}(\kappa)=-2\, {\cal L}_{\rm
scal}^{(1)(SD)}(\kappa)
\label{susy1l}
\end{eqnarray}
Note that this relation
holds for the renormalized effective Lagrangians, not
for the unrenormalized ones. It is due to a supersymmetry
of the self-dual background \cite{addvec}.
For real $\kappa$, this self-dual Lagrangian is real and 
has properties very similar
to the magnetic Lagrangian. 
Similarly, for purely imaginary $\kappa$ the self-dual
Lagrangian has an imaginary part, and provides a good analogue of the
electric field case. These cases will therefore be called `magnetic'
and `electric' in the following.
In particular, the imaginary part of the `electric' Lagrangian
has a Schwinger-type expansion

\begin{eqnarray}
{\rm Im} \left[{\cal
L}^{(1)(SD)}_{\rm scal}(i\kappa)\right]
=\frac{m^4}{(4\pi)^3}\frac{1}{\kappa^2}\sum_{k=1}^\infty
\left(\frac{2\pi\kappa}{k}+\frac{1}{k^2}\right)\, e^{-2\pi k \kappa}
\label{1lscimag}
\end{eqnarray}
Surprisingly, for the self-dual case even at two loops all parameter
integrals can be done in closed form, leading to the following
explicit formulas \cite{ds}:

\bear
{\cal L}_{\rm spin}^{(2)(SD)}(\kappa)
&=&
-2\alpha \,{m^4\over (4\pi)^3}\frac{1}{\kappa^2}\left[
3\xi^2 (\kappa)
-\xi'(\kappa)\right]
\non\\
{\cal L}_{\rm scal}^{(2)(SD)}(\kappa)
&=&
\alpha \,{m^4\over (4\pi)^3}\frac{1}{\kappa^2}\left[
{3\over 2}\xi^2 (\kappa)
-\xi'(\kappa)\right]
\non\\
\label{L2SD}
\ear\no
Here we have introduced the function $\xi$,

\bear
\xi(x)\equiv -x\Bigl(\psi(x)-\ln(x)+{1\over 2x}\Bigr)
\label{defxi}
\ear
and $\psi(\kappa) = {d\over d\kappa}{\rm ln}\Gamma(\kappa)$
is the digamma function. This function
has very simple expansions at zero as well as at
infinity,

\begin{eqnarray}
\psi(x)&\sim& -\frac{1}{x}-\gamma +\sum_{k=2}^\infty (-1)^k \zeta(k) x^{k-1}
\label{psiexpzero}\\
\psi(x)&\sim& \ln x-\frac{1}{2x}-\sum_{k=1}^\infty \frac{{\cal B}_{2k}}{2k
\,x^{2k}} 
\label{psiexpinf}
\end{eqnarray}
so that the formulas (\ref{L2SD}) directly yield
closed-form expressions for the coefficients of both 
the weak-field and strong-field expansions of the self-dual
Lagrangians.
Moreover, it is easy to obtain also a closed representation for
their imaginary parts which exist in the `electric' case:

\begin{eqnarray}
{\rm Im}\left[{\cal L}^{(2)}_{\rm scal}(i\kappa)\right]&=&
\alpha \pi
\frac{m^4}{(4\pi)^2}
\frac{1}{2\kappa}\sum_{k=1}^\infty 
\left[k-\frac{1}{2\pi\kappa}-
\frac{3\kappa}{2\pi} \sum_{l=1}^\infty
\frac{(-1)^l{\cal B}_{2l}}{2l{\kappa}^{2l}}\right]\, e^{-2\pi\kappa k}
\nonumber\\
{\rm Im}\left[{\cal L}^{(2)}_{\rm spin}(i\kappa)\right]&=&
-2\alpha \pi
\frac{m^4}{(4\pi)^2}
\frac{1}{2\kappa}\sum_{k=1}^\infty 
\left[k-\frac{1}{2\pi\kappa}-
\frac{3\kappa}{\pi} \sum_{l=1}^\infty
\frac{(-1)^l{\cal B}_{2l}}{2l{\kappa}^{2l}}\right]\, e^{-2\pi\kappa k}
\nonumber
\non\\
\label{fullimag2loopsd}
\end{eqnarray}
Note that these formulas display the same structure found by
Ritus and Lebedev for the physical electric case, 
(\ref{fullimag2l}),(\ref{expK}),
but that in the self-dual case we have simple closed formulas
for all the prefactors of the Schwinger exponentials. In 
(\ref{fullimag2loopsd}) we have written these prefactors
as series in ${1\over \kappa}$, which corresponds to the
$\beta$ - expansion (\ref{expK}). It is easy to see that this expansion
is asymptotic, rather than convergent, for all $k$.
For the scalar QED case, we have also verified \cite{ds} 
that the first ($k=1$) exponential in (\ref{fullimag2loopsd}),
including its complete prefactor, can be correctly reproduced
by an analysis of the asymptotic behaviour of the weak field
expansion coefficients and the application of the Borel
dispersion relations.
It is also immediately evident that, for any given value of $\kappa$,
the two-loop contribution to the prefactor of the $k$-th exponential
will dominate over the corresponding one-loop quantity in (\ref{1lscimag})
if $k$ is taken large enough.
Given the close similarity between the electric and `electric' cases
at the one-loop level it is natural to assume that these properties
hold also true for the electric case.

\section{The $N$ - photon amplitudes}

As is well-known, the Euler-Heisenberg Lagrangians (\ref{L1})
can be used to obtain the one-loop QED $N$ - photon amplitudes
in the low-energy approximation, i.e. for photon momenta $k_i$
such that $m^2$ is much larger than all $k_i\cdot k_j$ 's. After
specializing to a self-dual background it still contains the
information on the component of the $N$ - photon amplitude
with all helicities equal, say, all `+'. Using 
(\ref{L1}),(\ref{L2SD}),(\ref{psiexpinf}) 
together with the standard spinor helicity formalism
(see, e.g., \cite{lancerev}), one obtains for this low-energy limit
the following explicit formulas: At one loop,

\bear
\Gamma_{\rm scal}^{(1)(EH)}
[k_1,\varepsilon_1^+;\ldots ;k_N,\varepsilon_N^+]
&=&
\frac{(2e)^{N}}{(4\pi)^2m^{2N-4}}\,c_{\rm scal}^{(1)}
({\scriptstyle\frac{N}{2}})
\chi_N \non\\
c_{\rm scal}^{(1)}(n)&=& 
- \frac{{\cal B}_{2n}}{2n(2n-2)}
\non\\
\Gamma_{\rm spin}^{(1)(EH)}
[k_1,\varepsilon_1^+;\ldots ;k_N,\varepsilon_N^+]
&=&
-2\Gamma_{\rm scal}^{(1)(EH)}
[k_1,\varepsilon_1^+;\ldots ;k_N,\varepsilon_N^+]
\nonumber\\
\label{allplus1l}
\ear
and at two loops

\bear
\Gamma_{\rm scal}^{(2)(EH)}
[k_1,\varepsilon_1^+;\ldots ;k_N,\varepsilon_N^+]
&=&
\alpha\pi\frac{(2e)^{N}}{(4\pi)^2m^{2N-4}}\,c_{\rm scal}^{(2)}
({\scriptstyle{\frac{N}{2}}})
\chi_N 
\non\\
c^{(2)}_{\rm scal}(n)&=&
{1\over (2\pi)^2}\biggl\lbrace
\frac{2n-3}{2n-2}\,{\cal B}_{2n-2}
+\frac{3}{2}\sum_{k=1}^{n-1}
{{\cal B}_{2k}\over 2k}
{{\cal B}_{2n-2k}\over (2n-2k)}
\biggr\rbrace
\non\\
\Gamma_{\rm spin}^{(2)(EH)}
[k_1,\varepsilon_1^+;\ldots ;k_N,\varepsilon_N^+]
&=&
-2\alpha\pi\frac{(2e)^{N}}{(4\pi)^2m^{2N-4}}\,c_{\rm spin}^{(2)}
({\scriptstyle{\frac{N}{2}}})
\chi_N 
\non\\
c^{(2)}_{\rm spin}(n) &=&
{1\over (2\pi)^2}\biggl\lbrace
\frac{2n-3}{2n-2}\,{\cal B}_{2n-2}
+3\sum_{k=1}^{n-1}
{{\cal B}_{2k}\over 2k}
{{\cal B}_{2n-2k}\over (2n-2k)}
\biggr\rbrace
\non\\
\label{allplus2l}
\ear
In these formulas, the information on the external momenta is all
contained in the invariant $\chi_N$,

\bear
\chi_N 
&=& 
{({\scriptstyle \frac{N}{2}})!
\over 2^{N\over 2}}
\Bigl\lbrace
[12]^2[34]^2\cdots [(N-1)N]^2 + {\rm \,\, all \,\, permutations}
\Bigr\rbrace\non
\ear
Here $[ij]=\langle k_i^+\vert k_j^-\rangle$ denote
the basic spinor products \cite{lancerev}.

\section{Conclusions}

The existence of the simple closed-form expressions 
(\ref{L2SD}) for the self-dual two-loop
Euler-Heisenberg Lagrangians is surprising, and we do not know
of any comparable result in gauge theory. 
These results have allowed us to perform a more complete analysis of these
effective Lagrangians than is possible for other backgrounds
\cite{ds}, including the complete elucidation of the
structure of the imaginary parts. 
Moreover, we have used them to obtain explicit formulas for the
low-energy limits of the `all $+$' components of the
two-loop $N$ - photon amplitudes. 
Let us close with a remark on the behaviour of these amplitudes
in the limit where the number of photons becomes large. It is easy to
show that

\bear
\lim_{N\to\infty}\,\,
{\Gamma_{\rm spin,scal}^{(2)(EH)}
[k_1,\varepsilon_1^+;\ldots ;k_N,\varepsilon_N^+]
\over
\Gamma_{\rm spin,scal}^{(1)(EH)}
[k_1,\varepsilon_1^+;\ldots ;k_N,\varepsilon_N^+]}
&=&
\alpha\pi
\label{limNratio}
\ear
The Borel dispersion relations allow one to relate this quantity to
the $\kappa\to\infty$ limit of the corresponding ratio of the
imaginary parts (\ref{1lscimag}),(\ref{fullimag2loopsd}), and thus
to the factor $\alpha\pi$ appearing in the Lebedev-Ritus
exponentiation formula (\ref{shiftm}). 
Based on the expectation that this exponentiation occurs also in the
self-dual case, we conjecture that at loop order $l$, the weak-field expansion
will take the form 
\begin{eqnarray}
{\cal L}^{(l)(SD)}(\kappa)=\frac{(\alpha \pi)^{l-1}}{(l-1)!}\,
\frac{m^4}{(4\pi)^2}\sum_{n=2}^\infty c_n^{(l)}\, \frac{1}{\kappa^{2n}}
\label{conj}
\end{eqnarray}
where for each $l$, the expansion coefficients $c_n^{(l)}$ have the same {\it
leading} large $n$ growth rate as $c_n^{(1)}$. This would imply that

\bear
\lim_{N\to\infty}\,\,
{\Gamma_{\rm spin,scal}^{(l)(EH)}
[k_1,\varepsilon_1^+;\ldots ;k_N,\varepsilon_N^+]
\over
\Gamma_{\rm spin,scal}^{(1)(EH)}
[k_1,\varepsilon_1^+;\ldots ;k_N,\varepsilon_N^+]}
&=&
{(\alpha\pi)^{l-1}\over (l-1)!}
\label{limNratiogen}
\ear
It would be very interesting to verify this identity at the three-loop level.


\begin{thebibliography}{9}
\bibitem{leutwyler} H. Leutwyler, 
Phys. Lett. {\bf B 96} (1980) 154; Nucl. Phys. {\bf B 179} (1981) 129.
\bibitem{ward} R. S. Ward, Phil. Trans. Roy. Soc. Lond. {\bf A315} (1985) 451.
\bibitem{dufish} M.J. Duff and C.J. Isham, Phys. Lett. {\bf 86B}
(1979) 157; Nucl. Phys. {\bf B 162} (1980) 271.
\bibitem{allplustree} R. Gastmans and T. T. Wu, {\it The 
ubiquitous photon: helicity method for QED and QCD}, (Oxford Univerity Press,
New York, 1990).
\bibitem{allplusloop} G. Mahlon, Phys. Rev. D {\bf 49} (1994) 2197, 
hep-ph/9311213;
Z. Bern, L.J. Dixon, D.A. Kosower, 
Ann. Rev. Nucl. Part. Sci. {\bf 46} (1996) 109,
hep-ph/9602280; 
Z. Bern, A. De Freitas,
L. Dixon, A. Ghinculov and H. L. Wong, 
JHEP {\bf 0111} (2001) 031, hep-ph/0109079; 
Z.~Bern, A.~De Freitas and L.~Dixon,
JHEP {\bf 0203}, 018 (2002), hep-ph/0201161.
\bibitem{ds}
G.~V.~Dunne and C.~Schubert,
Phys.\ Lett. {\bf B 526}, (2002) 55, hep-th/0111134;
JHEP {\bf 0208} (2002) 053, hep-th/0205004;
JHEP {\bf 0206} (2002) 042, hep-th/0205005.
\bibitem{eh}
W. Heisenberg and H. Euler, Z. Phys. {\bf 98} (1936) 714.
\bibitem{schwinger}
J. Schwinger, Phys. Rev. {\bf 82} (1951) 664.
\bibitem{ritus}
V. I. Ritus, Zh. Eksp. Teor.
Fiz {\bf 69} (1975) 1517 [Sov. Phys. JETP {\bf 42} (1975) 774];
Zh. Eksp. Teor. Fiz {\bf 73} (1977) 807
[Sov. Phys. JETP {\bf 46} (1977) 423].
\bibitem{ditreu}
W.~Dittrich and M.~Reuter,
{\it Effective Lagrangians In Quantum Electrodynamics},
Springer Lect.\ Notes Phys.\  {\bf 220}  (1985).
\bibitem{frss}
M.~Reuter, M.~G.~Schmidt and C.~Schubert,
Annals Phys.\  {\bf 259}, 313 (1997), hep-th/9610191;
D.~ Fliegner, M.~ Reuter, M.~ G.~ Schmidt and C. ~Schubert, 
Teor. Mat. Fiz. {\bf 113} (1997) 289
[Theor. Math. Phys. {\bf 113} (1997) 1442], hep-th/9704194;
B. K{\"o}rs, M.G. Schmidt, 
Eur. Phys. J. {\bf C 6} (1999) 175, hep-th/9803144.
\bibitem{berkos}
Z. Bern and D. A. Kosower,
Nucl. Phys. {\bf B362} (1991) 389;
Nucl. Phys. {\bf B379} (1992) 451.
\bibitem{strassler}
M.J. Strassler, Nucl. Phys. {\bf B385} (1992) 145.
\bibitem{ss1}
M. G. Schmidt and C. Schubert, Phys. Lett. {\bf B318}
(1993) 438, hep-th/9309055.
\bibitem{review} C. Schubert, Phys. Rep. {\bf 355} (2001) 73, hep-th/0101036.
\bibitem{dunsch}
G. V. Dunne and C. Schubert, Nucl. Phys.
{\bf B 564} (2000) 591, hep-th/9907190. 
\bibitem{lebrit}
S. L. Lebedev, V. I. Ritus, 
Zh. Eksp. Teor. Fiz. {\bf 86} (1984) 408 [JETP {\bf 59} (1984) 237].
\bibitem{afalma}
I. K. Affleck, O. Alvarez, N.S. Manton,
Nucl. Phys. {\bf B 197} (1982) 509.
\bibitem{addvec}
A. D'Adda and P. Di Vecchia, Phys. Lett. {\bf 73 B} (1978) 162;
L.S. Brown and C. Lee, Phys. Rev. {\bf D 18} (1978) 2180.
\bibitem{lancerev}
L. Dixon, TASI Lectures, Boulder TASI 95, 539, hep-ph/9601359.

\end{thebibliography}
\end{document}